\documentclass[aps,pra,twocolumn,superscriptaddress,longbibliography]{revtex4-1}%
\usepackage{graphicx}
\usepackage{float}
\usepackage{dcolumn}
\usepackage{bm,color}
\usepackage{amsmath,amssymb,dsfont,amstext,amsfonts}
\usepackage{extarrows}
\usepackage[colorlinks=true,linkcolor=blue,urlcolor=blue,citecolor=blue]{hyperref}
\usepackage{xcolor}
\usepackage{wasysym}
\usepackage{mathtools}
\usepackage{bbold}

\usepackage[normalem]{ulem} 
\usepackage{color}

\newcommand{\parag}[1]{\vspace{0.3em}{\it #1.}}

\def\ket#1{\mathinner{|{#1}\rangle}}

\def\dd{\mathrm{d}}

\def\bs#1{\boldsymbol{#1}}

\begin{document}
\title{Topological Euler class as a dynamical observable in optical lattices}
\author{F. Nur \"{U}nal}
\thanks{Contributed equally. Refer for correspondence to \href{mailto:fnu20@cam.ac.uk}{fnu20@cam.ac.uk} and \href{mailto:rjs269@cam.ac.uk}{rjs269@cam.ac.uk}}
\affiliation{TCM Group, Cavendish Laboratory, University of Cambridge, J.J.~Thomson Avenue, Cambridge CB3 0HE, United Kingdom}
\author{Adrien Bouhon}
\affiliation{Nordic Institute for Theoretical Physics (NORDITA), Stockholm, Sweden}
\affiliation{Department of Physics and Astronomy, Uppsala University, Box 516, SE-751 21 Uppsala, Sweden}
\author{Robert-Jan Slager}
\thanks{Contributed equally. Refer for correspondence to \href{mailto:fnu20@cam.ac.uk}{fnu20@cam.ac.uk} and \href{mailto:rjs269@cam.ac.uk}{rjs269@cam.ac.uk}}
\affiliation{TCM Group, Cavendish Laboratory, University of Cambridge, J.J.~Thomson Avenue, Cambridge CB3 0HE, United Kingdom}

\date{\today}
\begin{abstract}
The last years have witnessed rapid progress in the topological characterization of out-of-equilibrium systems. 
 We report on robust signatures of a new type of topology~--the Euler class--~in such a dynamical setting.
The enigmatic invariant~$(\xi)$ falls outside conventional symmetry-eigenvalue indicated phases and, in simplest incarnation, is described by triples of bands that comprise a gapless pair featuring $2\xi$ stable band nodes, and a gapped band. These nodes host non-Abelian charges and can be further  undone by converting their charge upon intricate braiding mechanisms, revealing that Euler class is a fragile topology.
We theoretically demonstrate that quenching with non-trivial Euler Hamiltonian results in stable monopole-antimonopole pairs, which in turn induce a linking of momentum-time trajectories under the first Hopf map, making the invariant experimentally observable. Detailing explicit tomography protocols in a variety of cold-atom setups, our results provide a basis for exploring new topologies and their interplay with crystalline symmetries in optical lattices 
beyond paradigmatic Chern insulators.
\end{abstract}
\maketitle

\parag{Introduction}--
Topological insulators (TI) are gapped quantum phases that have a topological nature by virtue of protecting symmetries. Following
time reversal symmetry (TRS) protected TIs \cite{Rmp1,Rmp2}, past years have seen remarkable progress in characterizing topological materials taking into account crystal symmetries \cite{Clas2,Clas1, codefects2, Nodal_chains, HolAlex_Bloch_Oscillations, ShiozakiSatoGomiK, Chenprb2012, Codefects1, Wi2}. Using combinatorial arguments that map out classes of band structures in momentum space \cite{Clas3}, recent schemes formulated diagnosis criteria
upon comparing which of these combinations correspond to atomic configurations, defining topology relative to this subset \cite{Clas4, Clas5}. These pursuits also revealed fragile invariants that can be trivialized by gap closings with trivial bands, rather than involving those having opposite topological charge~\cite{Ft1, bouhon2019wilson, song2019fragile}. An archetypal invariant emerging from such studies, which goes beyond symmetry eigenvalue indicated phases \cite{Clas3, Clas4, Clas5} and relates to refined partitioning schemes~\cite{newpaperfragile}, is Euler class. It acts as the analogue of Chern number in systems
having $C_2\mathcal{T}$ [product of two-fold rotations and TRS] or $\mathcal{PT}$ [product of parity $\mathcal{P}$ and TRS] symmetry. All these aspects combined with band nodes featuring non-Abelian braiding properties make realizations of Euler class desirable \cite{Wu1273,BJY_nielsen, bouhon2019nonabelian}. However, concrete experimental signatures that would divulge the Euler invariant are still lacking.

On another note, ultracold atomic gases
have proven versatile platforms for exploring topological phenomena~\cite{Cooper19_RMP,Li16_Science_wilson,Aidelsburger15_NatPhys,Flaschner16_Sci,Asteria19_NatPhys, Jotzu14_Nat,Aidelsburger13_PRL,Miyake13_PRL,WangUnal_18_PRL,Wannierfloquet2020, Unal19_PRL, Wintersperger20_arx}. In particular, advances in periodic driving~\cite{Eckardt17_RMP} have called for expansion of these notions to out-of-equilibrium settings and new classification schemes~\cite{Kitagawa10_PRB,Rudner13_PRX,Roy17_PRB,Unal19_PRR,Dynamicalsynchronization2019}.
Development of novel quenching techniques, where a system is driven far from equilibrium by sudden changes in the Hamiltonian --most interestingly between topologically distinct regimes-- has not only  unearthed new connections between different topological invariants, but also provided a powerful experimental tool to detect topological quantities \cite{Wangchern_2017,Chen_ref_2020,Tarnowski19_NatCom,WeiPan18_PRL_hopfExp,YiPan19_arx_hopfTori,ZHANG18_sciBul,Caio15_PRL, Unal16_PRA,YangChen18_PRB,HuZhao20_PRL}. As versatile as they are, quench protocols have been employed only in the context of Chern insulators and mainly in two-band models. We here consider the dynamics of Euler class that requires a minimum of three bands (although the notion extends to many-band cases \cite{newpaperfragile, bouhon2019nonabelian}), following a quench between topologically distinct regimes. We demonstrate that non-trivial Euler class embodies stable monopole-antimonopole pair production, which provides for distinct experimental signatures in the linking of momentum-time trajectories upon appealing to a Hopf map. Accordingly, we device concrete experimental protocols to investigate this unexplored class of topology in cold-atom systems.

\parag{Chern number versus Euler class}--
In simplest form, Chern insulators can be described by a two-band Hamiltonian in two dimensions,
\begin{equation}\label{eq::2DChern}
H_{\mathcal{C}}(\mathbf{k})=\mathbf{d}(\mathbf{k})\cdot\boldsymbol{\sigma}+d_0(\bs{k})\sigma_0,
\end{equation}
for momentum $\bs{k}$ and Pauli matrices $\bs{\sigma}$. Upon spectral flattening the Hamiltonian (by replacing $\bs{d} \mapsto \bs{d}/||{\bs{d}}||$ and $d_0\mapsto 0$ without affecting the topology), the Chern number $\mathcal{C}$ coincides with the second homotopy group  of the sphere, $\pi_2(S^2)=\mathbf{Z}$. The invariant can readily be calculated via the Pontryagin skyrmion number \cite{Pontryagin41}
\begin{equation}\label{eq::pontrindex}
\mathcal{C}= \frac{1}{4\pi}\int_{BZ}\dd^2k~\mathbf{d}\cdot( \partial_{k_x} \mathbf{d} \,\times\, \partial_{k_y} \mathbf{d}),
\end{equation}
and can geometrically be interpreted as the covering of $S^2$ in terms of $\bs d$.

Similarly, Euler invariant $\xi$ finds a simple incarnation in three-band models. However, rather than a complex variant as in the Chern number, Euler class corresponds to a real characteristic form, which necessitates a protecting symmetry. One of the most rudimental is $C_2\mathcal{T}$ as almost all lattice geometries possess two-fold rotational symmetry, and thus will be assumed in the remainder. Consequently, the Hamiltonian can be recast into real symmetric matrix \cite{bouhon2019nonabelian}, having three eigenstates $E=\{\ket{u_1(\bs k)},\ket{u_2(\bs k)},\ket{u_3(\bs k)}\}$. The symmetry further dictates that the eigenstates form an orthonormal triad, a {\it dreibein}, satisfying $\bs n(\bs k)\equiv \bs u_3(\bs k)=\bs u_1(\bs k) \times\bs u_2(\bs k)$. In this basis, the associated spectral flattened form of the Euler Hamiltonian
reads, \cite{tomas,bouhon2019nonabelian}
\begin{equation}
H(\bs{k}) = 2\, \bs{n}(\bs{k})\cdot  \bs{n}(\bs{k})^\top -\mathbb{I}_3.\label{eqn::mainspecflat}
\end{equation}
The spectrum features two degenerate bands $\{u_1(\bs k),u_2(\bs k)\}$ with eigenvalue $-1$ and a `spectator' third band $\bs{n}(\bs{k})$ of energy $1$. Physically, the Euler invariant $\xi$ is encoded by winding of the degenerate two-band subspace, inducing a number of $2\xi$ band nodes, and traced by the third band due to the special structure of the {\it dreibein} (see Appendix~\textcolor{blue}{A} for details). The invariant then takes a particularly easy form \cite{bouhon2019nonabelian},
\begin{equation}
\xi= \frac{1}{2\pi}\int_{BZ}\dd^2k~\bs{n}\cdot \left(\partial_{k_x} \bs{n} \times \partial_{k_y} \bs{n} \right).\label{eq::Eulerdef}
\end{equation}
Surprisingly, the Euler class can thus geometrically be represented as a vector tracing the sphere, although there are no Chern bands.

Indeed, care has to be exercised in this sphere analogy. Vectors spanning the {\it dreibein} are {\it a priori} only defined up to a sign relating to the projective plane ${\bs R \bs P^2}$, as in a nematic \cite{Nissinen2016,Kamienrmp, Prx2016, volovik2018investigation, Beekman20171}. ${\bs R \bs P^2}$ features $\mathbf{Z}_2$-valued string charges and monopole charges, characterized by the first and second homotopy group (see Appendix~\textcolor{blue}{F}). Assuming absence of the former `weak invariants', the monopole charges are $\mathbf{Z}$-valued, although the sign is ambiguous, touching upon illustrious Alice dynamics \cite{AlicestringVolovik,Tiwari,Schwarz_alice}.
To construct Euler class, a consistent gauge, or handiness, has be to defined for $E$, making the analogy appropriate.

\parag{Model setting}--
The Hamiltonian can be generally parametrized as $H(\boldsymbol{k}) = R(\boldsymbol{k}) [-\mathbb{1}\oplus 1] R(\boldsymbol{k})^T$, where $R(\boldsymbol{k})$ accomplishes the desired winding via a geometric construction that makes use of a so-called Pl\"ucker embedding~\cite{newpaperfragile}. Sampling  over a grid set by the target lattice geometry, which we here take to be a square, renders the explicit hopping parameters specifying the Hamiltonian
\begin{equation}\label{eq::mainham}
H({\bs k})=\sum_jh_j({\bs k})\lambda_j,
\end{equation}
for the eight Gell-Mann matrices $\lambda_j$. In the following, we consider Euler class $\xi=2$ and $\xi=4$, the $h_j({\bs k})$ of which are given in Appendices~\textcolor{blue}{A,B} and~\textcolor{blue}{G}.

\begin{figure}
	\centering\includegraphics[width=1\linewidth]{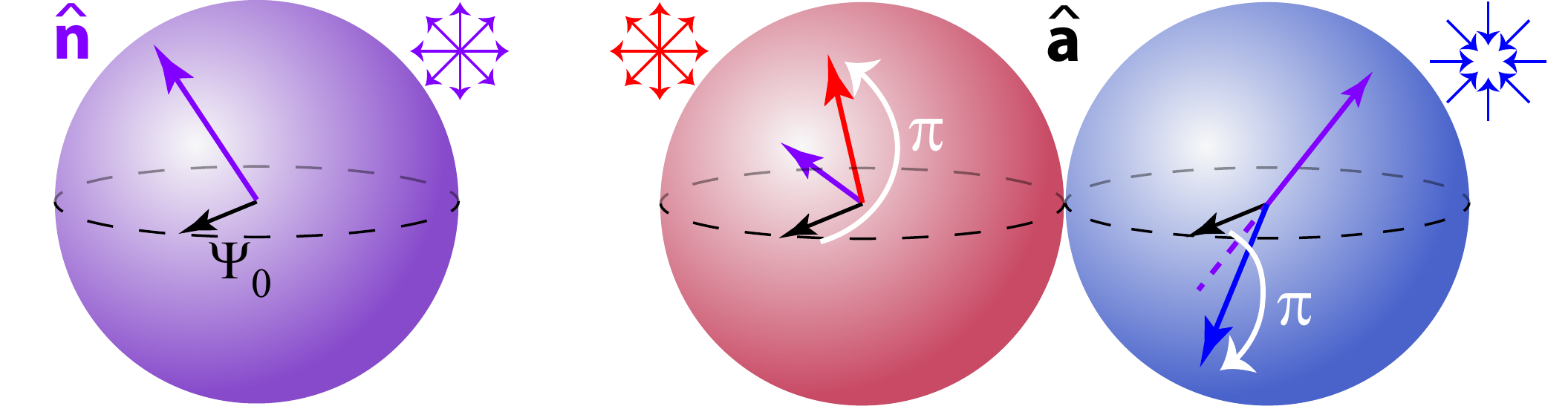}
	\caption{ Non-trivial Euler class can be related to a wrapping of the sphere $S^2$ by $\bs{n}(\bs{k})$ [left panel], inducing monopole charge, see Eq.\eqref{eq::Eulerdef}. The Hamiltonian induces a $\pi$-rotation around $\bs{n}(\bs{k})$. When acted upon initial state $\Psi_0$, the resulting vector $\bs{a}(\bs{k})=H(\bs{k})\Psi_0$, covers the sphere twice within the BZ, clockwise/anticlockwise depending on the orientation of $\bs{n}(\bs{k})$ with respect to $\Psi_0$, [red/blue vector on the right]. As a result, each time $\bs{n}(\bs{k})$ wraps $S^2$, $\bs{a}(\bs{k})$ wraps and unwraps, giving rise to a  monopole--antimonopole pair.
	}
	\label{fig1_spheres}
\end{figure}

\parag{Quench dynamics}-- We now turn to the quench dynamics.
For this purpose, we consider a trivial initial state $\Psi_0({\bs k})=(1,0,0)^{\top}$ and follow its time evolution, $\Psi({\bs k},t)= U(\bs{k},t)\Psi_0({\bs k})=e^{-itH(\bs{k})}\Psi_0({\bs k})$ resulting from a sudden change of the Hamiltonian to topologically non-trivial Euler form (we set $\hbar=1$). For simplicity, we spectrally flatten the quench Hamiltonian, which does not affect the topologically robust signatures presented.

The Euler Hamiltonian physically corresponds to a $\pi$-rotation around the vector $\bs{n}(\bs{k})$, as can be easily seen by analysing matrix components of Eq.\eqref{eqn::mainspecflat}. Since the initial state naturally corresponds to a normalized vector on $S^2$, this gives us the incentive to formulate the quench dynamics as rotations on this sphere. 
The evolving state characteristically traces a circle $S^1$ in time and, upon combining with the two-dimensional momentum space ($T^2$), forms a three-torus in $(k_x,k_y,t)$-space. As the weak invariants of this $T^3$ are zero, it corresponds to $S^3$ and gives the possibility of establishing a Hopf map from $S^3$ to $S^2$. We find that this is indeed the case. 
Using that the Hamiltonian squares to unity, the time evolution operator can be written in Rodrigues form, $\Psi(\bs{k},t)=[\cos{( t)}-i\sin{(t)}H(\bs{k})]\Psi_0(\bs{k})$. Although the complex $\Psi(\bs{k},t)$ is not tracing a Bloch sphere, we can project it back onto $S^2$ as $\bs{\hat{p}}=\Psi^{\dagger}(\bs{k},t)\bs{\mu}\Psi(\bs{k},t)$, where
\begin{equation} \label{eq::Mu_definitions}
 \mu_x\!=\!\!
 \begin{pmatrix}
0 &  i & 1\\
-i & 0 & 0\\
1 & 0 & 0
\end{pmatrix}\!\!,
\;
\mu_y\!=\!\!
 \begin{pmatrix}
0 &  1 & -i\\
1 & 0 & 0\\
i & 0 & 0
\end{pmatrix}\!\!,
\;
\mu_z\!=\!\!
\begin{pmatrix}
1 &  0 & 0\\
0 & -1 & 0\\
0 & 0 & -1
\end{pmatrix}\!,
\end{equation}
defining the time circle $t\in[0,2\pi/\Delta)$ for energy gap $\Delta=2$ of the flattened Hamiltonian.

The above construction can be best motivated upon appealing to the quaternion description of the Hopf map. Quaternions, written as $q=x_0+x_1{ i}+x_2{ j}+x_3{ k}, x_i \in \mathbf{R}^4$,   extend complex numbers, with units satisfying ${ij}=k, {jk}=i, {ki}=j$ and ${i}^2={j}^2={k}^2=-1$. Vectors in $\mathbf{R}^3$ can be represented as pure quaternions with real part $x_0=0$, using the units ${i,j,k}$ as basis vectors. Moreover, quaternions having unit norm $\sqrt{x_0^2+x_1^2+x_2^2+x_3^2}=1$ are called {\it versors} and implement rotations on 3D vectors.
In particular, acting with versor $v=x_0+\mathbf{v}$ on vector $\bf{t}$ as
$R_v: {\bf t}\mapsto v{\bf{t}} v^{-1}$, where $v^{-1}$ is the inverse of $v$, implements a rotation around vector $\mathbf{v}$ by an angle $\theta=2\arccos{(x_0)}$. The rotation map can be rewritten in matrix form $R_v {\bs t}={\bf t'}$, with
\begin{eqnarray}\label{eq::mainrotationversorrep}
\scriptsize
&R_v=  \nonumber\\
&\!\!\!\!\begin{pmatrix}
x_0^2+x_1^2-x_2^2-x_3^2, \!\!&\!\! 2x_1x_2-2x_0x_3, \!\!&\!\! 2x_0x_2+2x_1x_3 \\
2x_1x_2+2x_0x_3, \!\!&\!\! x_0^2-x_1^2+x_2^2-x_3^2, \!\!&\!\! -2x_0x_1+2x_2x_3\\
-2x_0x_2+2x_1x_3, \!\!&\!\! 2x_0x_1+2x_2x_3,\! \!&\!\! x_0^2-x_1^2-x_2^2+x_3^2
\end{pmatrix}\!\!.  \nonumber
\end{eqnarray}
This description carries a deeper meaning as a representation of the Hopf map. Since versors have unit norm, they span a three-sphere $S^3\subset\mathbf{R}^4$. In addition, the rotation map preserves the norm and hence, acting on a normalized vector, results in another normalized vector on $S^2$. In other words, acting with $R_v$ on a unit norm vector (from the left or right) induces a map from $S^3$ to $S^2$, the renowned Hopf map. The inverse image is a circle, constituting the familiar fibre. A key insight is that the elements of $\Psi(t)$ can be related to a quaternion by associating $\{x_0,\bs{x}\}=\{\cos{t}, -\sin(t)H\Psi_0\}$, which then connects to $R_v$ through the $\bs{\mu}$-matrices. We refer to Appendix~\textcolor{blue}{C} for further elaboration on the Hopf parametrization.

\parag{Monopole-antimonopole pairs}--
We now analyze the physical consequences embodied by the Hopf construction. We identify ${\bs a ({\bs k})}=H({\bs k})\Psi_0({\bs k})$ as a $\pi$-rotation of the initial state around the vector ${\bs n ({\bs k})}=(n_1,n_2,n_2)$ as illustrated in Fig.\ref{fig1_spheres}.
For $\Psi_0({\bs k})=(1,0,0)^{\top}$, this results in ${\bs a ({\bs k})}=(2n_1^2-1,2n_1n_2,2n_1n_3)^{\top}$ through Eq.\eqref{eqn::mainspecflat}.
Where ${\bs n ({\bs k})}$ wraps the sphere once [left-purple sphere in Fig.\ref{fig1_spheres}], the vector ${\bs a ({\bs k})}$ covers the sphere twice within the Brillouin zone. For $\bs{k}$-values where ${\bs n ({\bs k})}$ lies within the same hemisphere with $\Psi_0$ (i.e.~$+\hat{x}$-hemisphere), the orientation of the wrapping of ${\bs a ({\bs k})}$ is same with ${\bs n ({\bs k})}$ --clockwise, encapsulating a monopole [red vector-sphere]. However, when ${\bs n ({\bs k})}$ and $\Psi_0$ lie on opposite hemispheres, ${\bs a ({\bs k})}$ covers the sphere anticlockwise [blue sphere], corresponding to an antimonopole.

Analytically, this can be seen by writing ${\bs n}=(\cos\alpha,\sin\alpha\cos\beta,\sin\alpha\sin\beta)$ in spherical coordinates, where the polar angle $\alpha$ is defined with respect to $+\hat{x}$-axis, and the azimuthal angle $\beta$ from $+\hat{y}$-axis for simplicity. Correspondingly, the vector ${\bs a}=(\cos2\alpha,\sin2\alpha\cos\beta,\sin2\alpha\sin\beta)$ then features doubling of the polar angle as compared to ${\bs n}$, hence covering the sphere twice. For $\alpha\in[0,\pi/2)$ this is clockwise, and anticlockwise for $\alpha\in[\pi/2,\pi)$ as $\sin2\alpha$ changes sign. Alternatively, considering the stereo-graphic representation of ${\bs a ({\bs k})}$ results in the same conclusion. We emphasize that this monopole--antimonopole pair cannot annihilate each other through recombination and are topologically stable, since the patches of the Brillouin zone (BZ) hosting them belong to the clockwise and anticlockwise cover and thus are naturally separated, akin to the double cover of the BZ for Chern insulators having invariant 2.

Following the quench, the time-evolving state inherits this monopole--antimonopole structure through
\begin{equation}\label{eq::timeevolmain}
\Psi({\bs k},t)=\cos(t)\Psi_0({\bs k})-i\sin(t){\bs a({\bs k})},
\end{equation}
which will be also imprinted on the linking number. Under the Hopf construction, each point in the $(k_x,k_y,t)$-space maps to a vector $\bs{\hat{p}}$ on the Bloch sphere via Eqs.~\eqref{eq::timeevolmain} and \eqref{eq::Mu_definitions}. When the Hopf map is non-trivial, inverse images of {\it any} two such vectors on $S^2$ manifest linking in $T^3$  \cite{Wilczek}, the number of which signifies the underlying invariant. Given the (anti-)monopole structure, a linking number on complementary patches of the BZ is anticipated.

\begin{figure}
	\centering\includegraphics[width=1\linewidth]{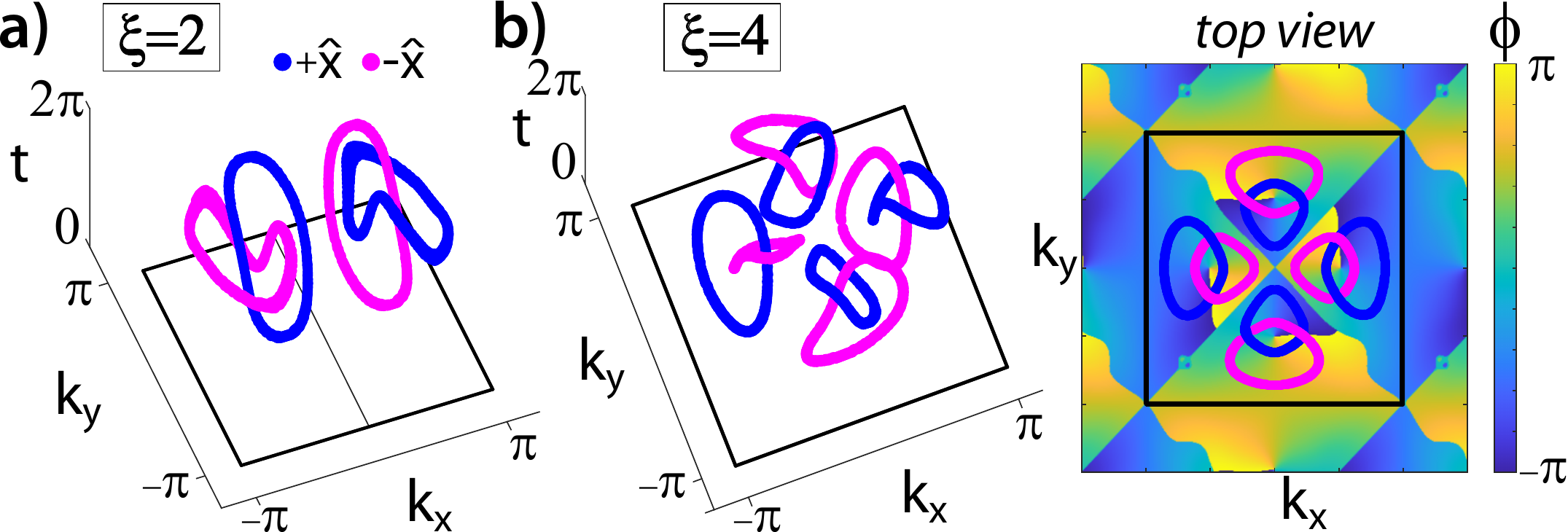}
	\caption{Hopf--linking structure of Euler class. a) For Euler number $\xi=2$, the inverse images of $\pm\hat{x}$ trace a link and anti-link on two patches in the BZ. The gap is rescaled so that the time circle runs through $t\in[0,2\pi)$.
	(b) $\xi=4$ with two pairs of oppositely oriented links. Four different patches of the BZ clearly visible in the top view of the azimuthal angle $\phi$ of the Bloch sphere.
	}
	\label{fig2_linking}
\end{figure}

To demonstrate this, we numerically evolve the initial state following a quench with a Hamiltonian having Euler invariant $\xi$. In Fig.\ref{fig2_linking}, we present the inverse images of two vectors $\bs{\hat{p}_1}=(1,0,0)$ and $\bs{\hat{p}_2}=(-1,0,0)$, i.e.~we mark each $(k_x,k_y,t)$-triple at which the state points along $\bs{\hat{p}_1}$ or $\bs{\hat{p}_2}$. In Fig.\ref{fig2_linking}a for $\xi=2$, the inverse images link twice with opposite sign in separate patches of the BZ, conforming the monopole--antimonopole pair. Similarly, when the Euler invariant is $\xi=4$ as in Fig.\ref{fig2_linking}b, there are four different regions attributed to two linkings of positive, and two of negative sign.

We quantify our findings upon appealing to the Hopf invariant $\mathcal{H}$. Here, it is convenient to use a complex two-vector basis, as it reinforces the quaternion analogy.
A quaternion can be rewritten as $q=z_1+z_2j$ by using two complex numbers $z_1=x_0+ix_1$ and $z_2=-x_2+ix_3$. Accordingly, a two-vector description of $\Psi({\bs k},t)$ may be established by identifying its first component (the one harboring the real part $x_0$) as $z_1({\bs k}, t)$, and combining the second and third components into a single function as $z_2({\bs k},t)=i\Psi_2({\bs k},t)+\Psi_3({\bs k},t)$. We thus define $\zeta({\bs k},t)=(z_1({\bs k},t),z_2({\bs k},t))^{\top}$. The Bloch vector is then reproduced via Pauli matrices amouting to the standard Hopf parametrization, $\bs{\hat{p}}=\zeta^{\dagger}({\bs k},t){\bs \sigma}\zeta({\bs k},t)$, which was notably employed
to reveal the Chern number \cite{Wangchern_2017}. After inserting the two-vector $\zeta({\bs k},t)$ in the Hopf invariant,
\begin{equation}\label{eq::Hopfinv}
\mathcal{H}=-\frac{1}{4\pi^2}\int \dd^2k\:dt\; \epsilon^{abc}\; \zeta^{\dag} \partial_a\zeta \partial_b\zeta^{\dag} \partial_c\zeta,
\end{equation}
the time integral decouples as in the Chern case \cite{Wangchern_2017,ChernHopf_Yu,Chang2018}, and we find that the Hopf invariant $\mathcal{H}$ reduces to the winding of $\mathbf{a}$,
\begin{equation}\label{eq::Hopfinv_a}
\mathcal{H}
=\frac{1}{4\pi}\int_{BZ}\dd^2k~\mathbf{a}\cdot( \partial_{k_x} \mathbf{a} \,\times\, \partial_{k_y} \mathbf{a}).
\end{equation}

Due to monopole--antimonopole pairs, $\mathcal{H}$ evidently attains zero value.
Nonetheless, as anticipated, restricting the domain of integration in Eq. \eqref{eq::Hopfinv} to the the left/right half of the BZ in Fig.\ref{fig2_linking}a, corresponding to a single wrapping of the sphere, we find $\mathcal{H}_{\pm}=\pm1$ for the patches.  
The Euler form corresponds to $\mathcal{H}'=\sum_{+}\mathcal{H}_{\alpha}-\sum_{-}\mathcal{H}_{\alpha}$, where $+(-)$ spans patches having (anti-)monopole charge $\mathcal{H}_{\alpha}$, distinguishing the monopole--antimonopole configuration for $\xi=2$ and $4$ in Fig.~\ref{fig2_linking}. 
Moreover, quenches from non-trivial to trivial Euler class can be shown to imprint linking proportional to the change in the invariant.

We stress that the monopole description of $\pi$-rotated vector ${\bs a}({\bs k})$ is independent of initial trivial state $\Psi_0({\bs k})$. As a result, although the specific parametrization of the Hopf map and the linking pattern changes, the topological content of the Hopf construction is robust and can be worked out for any state as detailed in Appendix~\textcolor{blue}{D}.

\begin{figure}
	\centering\includegraphics[width=1\linewidth]{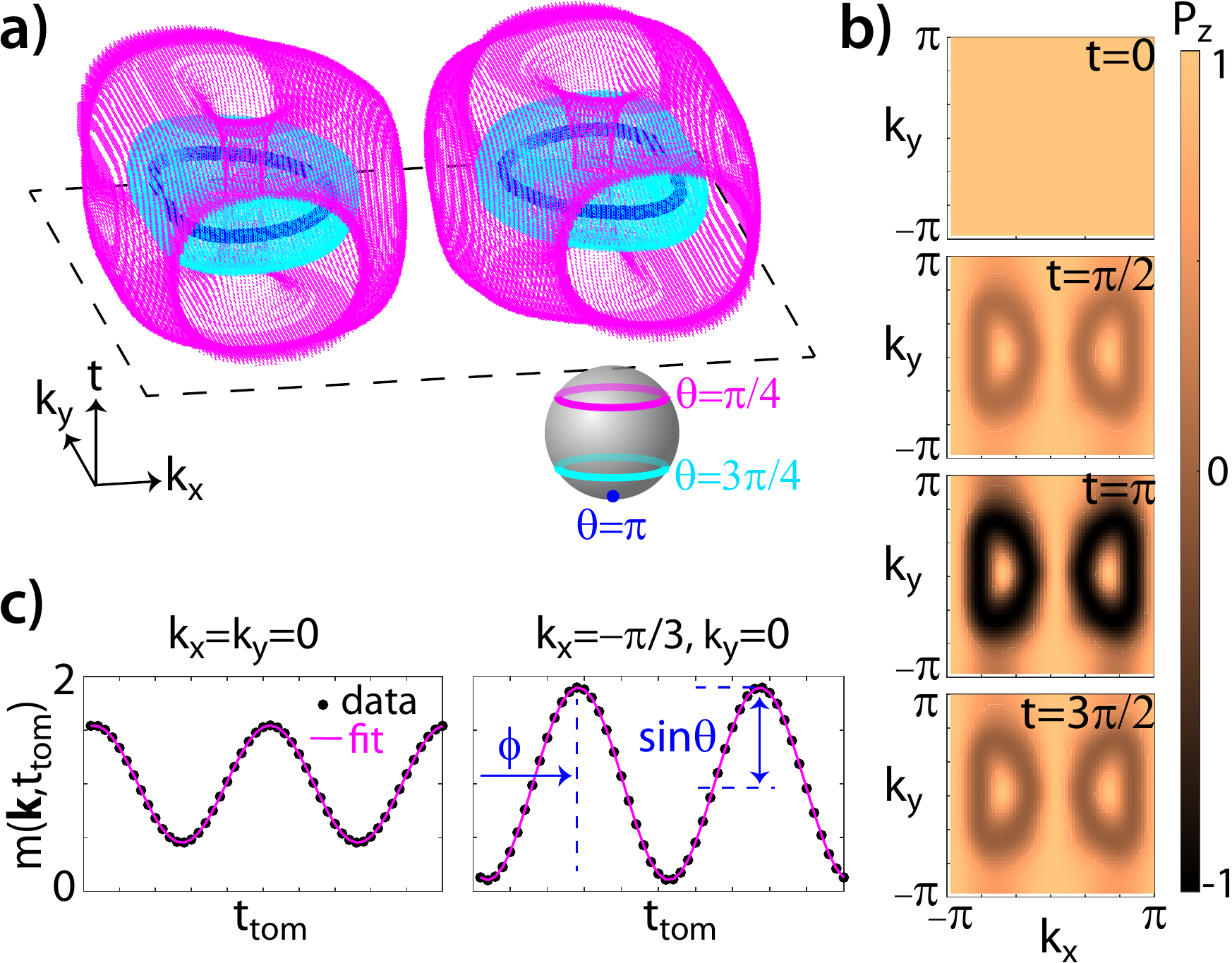}
	\caption{Proposed experimental schemes. 
	a) Inverse image of a circle on $S^2$ corresponds to a torus in $(k_x,k_y,t)$-space, which can be measured as polarization $P_z=\cos(\theta)$. For $\xi=2$, different latitudinal circles ($\theta$) form two nested Hopf tori within the BZ (dashed line) due to the monopole--antimonopole signature of Euler class. b) Snapshots of $P_z$ at different quench times $t\in[0,2\pi)$. c) The linking of the inverse images can be also detected via a state tomography as detailed in the text.
	Phase and amplitude of the oscillations in the
	momentum density coming from TOF reveal the Bloch sphere angles, here illustrated for two representative $\bs{k}$-values and $t=\pi$.
	}
	\label{fig3_experiment}
\end{figure}

\parag{Experimental realization and measurement protocols}--
The Euler model Eq.\eqref{eq::mainham} can be adopted to different lattice geometries (e.g. square, Kagome), and optimized for e.g.~number of hopping terms (see Appendix~\textcolor{blue}{A} and~\cite{newpaperfragile}), giving considerable flexibility for experimental realization. \textcolor{orange}{To underpin this, we present another model in real space with fewer number of tunneling elements in Appendix}~\textcolor{blue}{B}.
The pseudospin flavor of the model can be most easily encoded as different internal states in optical lattices~\cite{WeiPan18_PRL_hopfExp,YiPan19_arx_hopfTori}, where the versatility of optical flux lattices might be also used to engineer non-trivial models in momentum space~\cite{Cooper11_PRL_fluxlatt}.

When the pseudospin structure is induced by using different hyperfine states ($|A\rangle,|B\rangle,|C\rangle$), the quench from a spin-polarized trivial initial state ($\Psi_0=|A\rangle$) to a non-trivial Euler form initiates Raman-induced oscillations in spin polarization that can be measured locally for each $\bs{k}$~\cite{WeiPan18_PRL_hopfExp,YiPan19_arx_hopfTori}.~The $\hat{z}$-component of the Bloch vector defined via Eq.\eqref{eq::Mu_definitions} is then given by $P_z(\bs{k},t)=(N_A-N_B-N_C)/(N_A+N_B+N_C)$, where $N_{\gamma}$ is the number of particles detected in spin state $\gamma=A,B,C$. Inverse images of circles defined as $P_z(\bs{k},t)=\cos\theta$ on the Bloch sphere for polar angle $\theta$ traces closed surfaces in $T^3$. We illustrate these nested (non-trivial) Hopf tori for $\xi=2$ resulting from monopole--antimonopole structure in Fig.~\ref{fig3_experiment}a-b.

Alternatively, in optical lattices with sublattice degrees of freedom, we adopt the state tomography techniques~\cite{Hauke14_PRL,Alba11_PRL}, which has been restricted to two bands so far for detecting the linking structure. The momentum distribution of the quenched state can be measured in time-of-flight (TOF) as
$m(\bs{k})=\tilde{w}(\bs{k})|(\langle A|+\langle B|+\langle C|)|\Psi(\bs{k},t)\rangle|^2$, where $\tilde{w}(\bs{k})$ is the Fourier transform of the Wannier function~\cite{Hauke14_PRL,Flaschner16_Sci}. However, as it does not correspond to the Bloch sphere, $m(\bs{k})$ on its own does not capture the winding of the Euler class on $S^2$. By analyzing the two-vector $\zeta(\bs{k},t)$ description of the Hopf map, one can see that the north and south poles of the sphere can be associated to $|A\rangle$ and $|D\rangle=i|B\rangle+|C\rangle$, so that $\zeta(\bs{k},t)=\sin(\theta/2)|A\rangle+\cos(\theta/2)e^{i\phi}|D\rangle$ in spherical coordinates. To access the Bloch sphere in TOF, we apply a $\pi/2$-pulse to the state with respect to the sublattice $B$. Namely, we evolve $\Psi(\bs{k},t)$ with the flat-band Hamiltonian $H_{pulse}=(\nu/2)\:\text{diag}(1,-1,1)$, for a time $\pi/2\nu$ given by the energy difference $\nu$, so that the second component of the state acquires a phase $e^{i\pi/2}$; i.e.~$\Psi'(\bs{k},t)=(\Psi_1(\bs{k},t),i\Psi_2(\bs{k},t),\Psi_3(\bs{k},t))^{\top}$, directly corresponding to $\zeta(\bs{k},t)$. Upon further projecting on flat bands for tomography, $H_{tom}=(\omega/2)\mu_z$, the state $\Psi'(\bs{k},t)$ starts precessing around the $\hat{z}$-axis with frequency $\omega$, adding a phase $e^{-i\omega t_{tom}}$ to $\Psi_1(\bs{k},t)$ within time $t_{tom}$. As such, the momentum distribution after TOF reveals the Bloch vector via $m(\bs{k},t_{tom})=\tilde{w}(\bs{k})(1+\sin\theta_{\bs{k}}\cos(\phi_{\bs{k}}+\omega t_{tom}))$, where we fit with a cosine to extract the amplitude and the phase~\cite{Hauke14_PRL,Flaschner16_Sci}. We display these oscillations in $m(\bs{k},t_{tom})$ in Fig.~\ref{fig3_experiment}c, and indeed retrieve the linking due to monopole--antimonopole structure by reconstructing
Bloch angles ($\theta,\phi$)--see Appendix~\textcolor{blue}{E} for explicit results.

\parag{Conclusions}-- We have demonstrated that quench dynamics of Euler class  naturally embodies a Hopf construction, where the non-trivial winding of Euler Hamiltonian on $S^2$ generates stable monopole--antimonople pairs living in separate BZ patches.
Upon appealing to quaternions, we prove that the Hopf invariant and linking number captures this monopole--antimonople signature.
We show that the nested Hopf tori and the non-trivial linking in $(k_x,k_y,t)$-space can be detected in spin-resolved measurements in momentum space, or via a modified state tomography technique which we expand to the three-band Euler model. Our work opens up avenues for the exploration of new crystalline and exotic fragile topologies that have attracted much interest in the last couple of years, in the versatile setting of ultracold quantum gases of three-band systems and beyond.

\begin{acknowledgments}
	{\it Acknowledgments --}
We thank Tomas Bzdusek and Nigel Cooper for fruitful discussions. F.N.\"{U}.~acknowledges funding from EPSRC Grant No.~EP/P009565/1 and the Royal Society Newton International Fellowship. R.-J.~S.~acknowledges funding from the Marie Sk{\l}odowska-Curie programme under EC Grant No. 842901 and the Winton programme as well as Trinity College at the University of Cambridge.
\end{acknowledgments}

\bibliography{references}
\appendix
\section{Hamiltonians for three band models having non-trivial Euler class}\label{sec:models}
We here briefly comment on the explicit models of the main text that were derived using the techniques of Ref. \cite{newpaperfragile}. These include systems having Euler class $\xi=2,4$ and different spectral orderings of the bands. Finally, we also illustrate the derived Chern models induced by the non-degenerate subspace. 

\parag{Model Hamiltonians}-- Chern insulator models find their simplest incarnation in two-band systems of the form 
\begin{equation}\label{appeq::2band}
H_{\mathcal{C}}=\bf{ d}(\bf{k}) \cdot \boldsymbol{\sigma}+d_0(\bs{k})\sigma_0,   
\end{equation}
on the bases of Pauli matrices ${\bs \sigma}$. Hamiltonians exhibiting Euler class have a similar tractable form in terms of real symmetric three-band models, that have two degenerate bands featuring a number of $2\xi$ of isolated band nodes, and a third separated band. A priori, one may expect that separating the system into a two-band $\{\ket{u_1(\bs k)},\ket{u_2(\bs k)}\}$ and one-band subspace $\{\ket{u_3(\bs k)}\}$ can give nontrivial behavior as the stable homotopy classes of Wilson flows, characterized by the first homotopy group $\pi_1$, cover $S^2$ \cite{bouhon2019wilson,tomas}. The two subblocks then accordingly relate to $\pi_1(SO(2))=\mathbf{Z}$, conveying the gapless charges within the gap between these two bands, whereas $\pi_1(SO(1))=0$ relates to the third band. However, these guiding intuitions should be taken with caution. First of all, the orthonormal frame spanned by $E=\{\ket{u_1(\bs k)},\ket{u_2(\bs k)},\ket{u_3(\bs k)}\}$ is invariant under reversing the sign of each of the vectors. Hence, the space of Hamiltonians is actually given by the projective plane $S^2/\bs Z_2=\bs R \bs P^2$, coinciding with bi-axial nematic descriptions \cite{Kamienrmp,Prx2016}. Fortunately, as for the Euler classes, we are only interested in the orientable case of the vector bundle which {\it can} actually be related to the sphere \cite{bouhon2019nonabelian}. Secondly, the facts that all bands must sum up to a trivial charge and $\pi_1(SO(3))=\mathbf{Z}_2$ also suggest that the integer winding of the two band subspace has to be even, which is consistent with earlier findings that an odd Euler class would require a four-band model \cite{Ahn2018b}. Finally,  $\ket{u_3(\bs k)}$  {\it is} related to the other two eigenstates as $\ket{u_3(\bs k)}=\ket{u_1(\bs k)}\times\ket{u_2(\bs k)}\equiv\bs n(\bs k)$. Although these hints are merely a motivation, the spectrally flattened form for such real three-band systems can indeed be shown to be  \cite{bouhon2019nonabelian, newpaperfragile,tomas}, 
\begin{equation} \label{eq::specflat;app}
H(\bs{k}) = 2\, \bs{n}(\bs{k})\cdot  \bs{n}(\bs{k})^\top -\mathbb{I}_3.
\end{equation}

With these insights it is possible to construct Euler Hamiltonians systematically using a geometric construction  on arbitrary lattice geometries \cite{newpaperfragile}. In a nutshell, the idea is that the winding can be encoded via a so-called Pl\"ucker embedding.
That is, via a pullback map to coordinates paramterizing the sphere, the winding can be formulated in terms of rotation matrix $R({\bs k})$. 
Concretely, this means that we can obtain Hamiltonians with Euler class $\xi$ as 
\begin{equation}\label{eq::construct}
H(\boldsymbol{k}) = R(\boldsymbol{k}) [-\mathbb{1}\oplus 1] R(\boldsymbol{k})^T,    
\end{equation}
where $R({\bs k})$ implements a $\xi$-times winding of the sphere and we take the flattened energies of degenerate/third band subspace to be $-/+1$.

From $H(\bs{k})$ we can then get an explicit tight-binding model upon sampling over a grid $\Lambda^*$ set by the lattice, e.g a square lattice. Applying an inverse discrete Fourier transform then results in the hopping paramaters $t_{\alpha\beta}(\boldsymbol{R}_j-\boldsymbol{0}) = \mathcal{F}^{-1}[ \{H_{\alpha\beta}(\boldsymbol{k}_m)\}_{m\in \Lambda^*}](\boldsymbol{R}_j-\boldsymbol{0}) $ around a specific site. Formally this has to be executed over the whole Bravais lattice $\boldsymbol{R}_j\in \{l_1 \boldsymbol{a}_1 + l_2 \boldsymbol{a}_2\}_{l_1,l_2\in\mathbf{Z}}$ (here spanned by the primitive lattice vectors of the square lattice $\boldsymbol{a}_1 = a \hat{x}$, $\boldsymbol{a}_2 = a \hat{y}$), but due to fast decay of the Fourier envelope we can get simple models by truncating this process, meaning that we consider a specific number of neighbors $\boldsymbol{R}_j\in \{ l_1 \boldsymbol{a}_1 + l_2 \boldsymbol{a}_2 \}_{ 0\leq \vert l_1\vert,\vert l_2\vert \leq N}$ for tunneling.

For the $\xi=2$ case, we restrict tunneling to $N=2$ neighbors. The three-band Hamiltonian then takes the form, 
\begin{equation} \label{eq::App:mainham}
H({\bs k})=h_j({\bs k})\lambda_j,     
\end{equation}
in terms of the five real Gell-Mann matrices $\lambda_j$, where the sum over $j$ is implied. The specific form of the $h_j({\bs k})$ then read
\begin{equation}\label{eq::modelparameters}
h_j({\bs k})=\sum_{l_1,l_2}t_j(l_1,l_2)e^{i (l_1 k_x+l_2 k_y)},     
\end{equation}
where $l_1, l_2$ are integers running from $-N$ to $N$ and $t_j(l_1,l_2)$ defines a matrix element of hopping parameters obtained by the truncation procedure defined above. We have specified these eight $5\times5$-matrices for completeness in Appendix \ref{App:matrices Hamiltonian}.

Similarly, it is straightforward to obtain models having Euler class $\xi=4$, where we truncate to $N=3$ neighbors to accommodate the higher winding. The hopping matrices of this higher winding are also specified in Appendix  \ref{App:matrices Hamiltonian}. Finally, the construction is also flexible to shift the third band to the bottom of the spectrum, corresponding to the flattened form, 
\begin{equation}
H(\bs{k}) =\mathbb{I}_3- 2\, \bs{n}(\bs{k})\cdot  \bs{n}(\bs{k})^\top, 
\end{equation}
by acting with $R$ as $H(\boldsymbol{k}) = R(\boldsymbol{k}) [-1\oplus\mathbb{1}] R(\boldsymbol{k})^T$. We refer 
to these systems as``inverted".

\begin{figure}
	\centering\includegraphics[width=.85\linewidth]{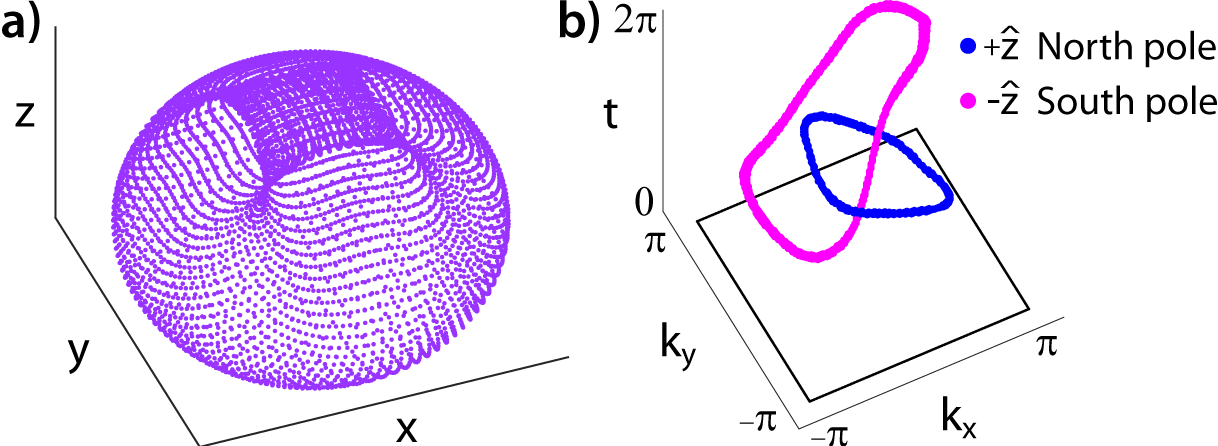}
	\caption{ Topology of $n(\bs{k})$ for $\xi=2$. a) The vector $n(\bs{k})$ covers $S^2$ once. b) Inverse images of the poles link once when a trivial state quenched to evolve under $n(\bs{k})$.	}
	\label{fig:App1_n(k)}
\end{figure}

\parag{Derived Chern model}--
As stated in the main text, the class of Euler Hamiltonians that are studied here  directly induce Chern insulator models upon promoting the normalized ${\bs n(\bs k)}=\ket{u_3(\bs k)}$-eigenstate to $d({\bs k})$ in Eq. \eqref{appeq::2band}. We stress however once more that there are no Chern numbers in the Euler system and this mereley induces a new Chern Hamiltonian.
Since $n(\bs{k})$ is always normalized and real (due to the symmetries), it can be treated as a Bloch vector that corresponds to a flattened Hamiltonian as $H'(\bs{k})=n(\bs{k})\cdot \bs{\sigma}$. We consider the Euler Hamiltonian with $\xi=2$ and display the complete coverage of two-sphere by $n(\bs{k})$ in Fig.~\ref{fig:App1_n(k)}a. Accordingly, when we quench a trivial initial state ($\Psi_0(\bs{k})=\frac{1}{\sqrt{10}}(1,2)^\top$) to suddenly evolve with $H'(\bs{k})$, we obtain linking number (Hopf invariant) 1 in Fig.~\ref{fig:App1_n(k)}b.

\section{Alternative simple model in square lattice}
\begin{figure}
	\centering\includegraphics[width=.7\linewidth]{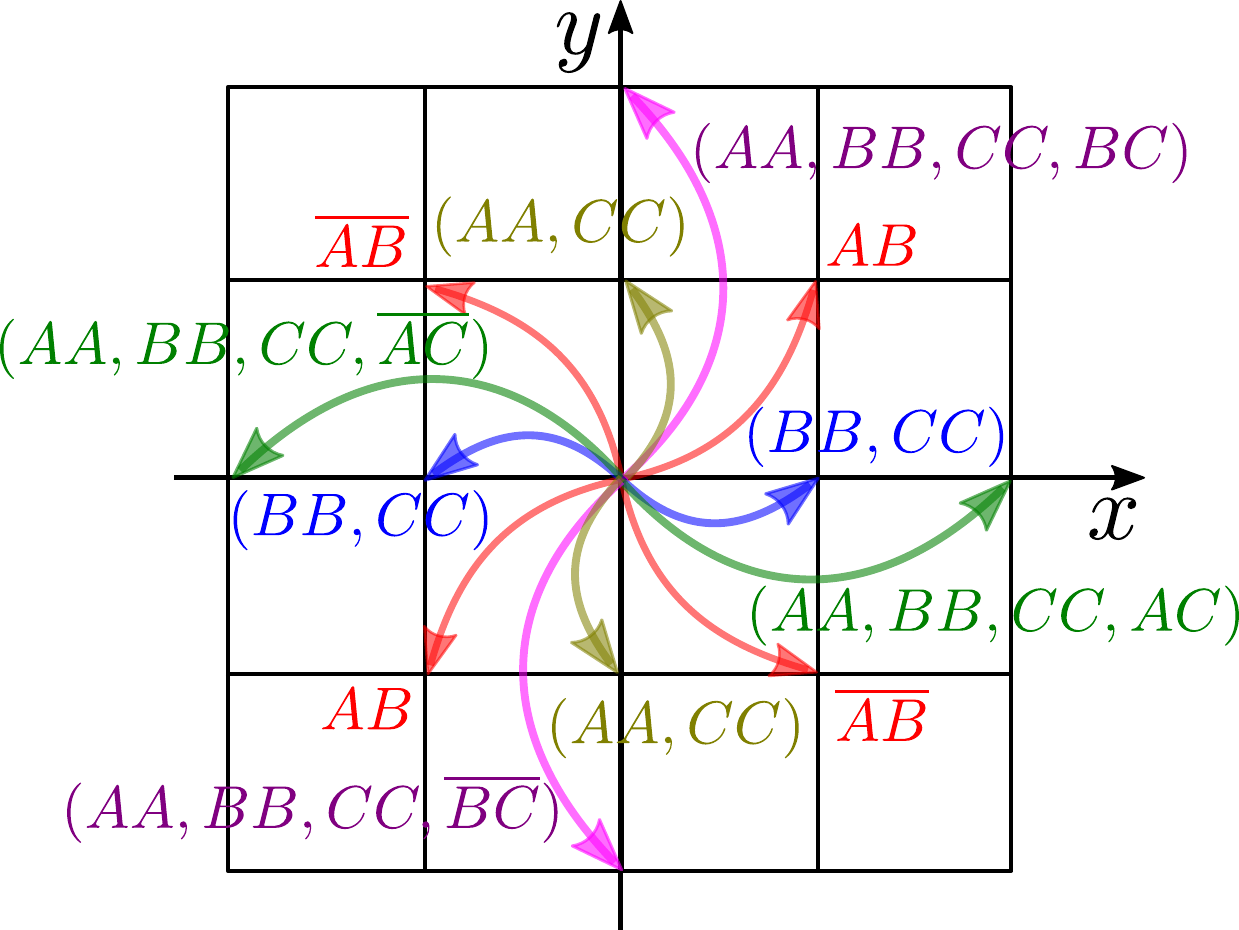}
	\caption{Real space hopping parameters defining the model~\eqref{eq::alternativemodel}.
	Each site has three orbitals $A,B,C$ and hopping terms $t^{(l_1,l_2)}_{\alpha\beta}$, i.e.~from orbital $\alpha$ to $\beta$ along the vector $(l_1,l_2)=l_1 \hat{x} + l_2\hat{y}$, are indicated as $\alpha\beta$, and $\overline{\alpha\beta}$ for $-t^{(l_1,l_2)}_{\alpha\beta}$. }
	\label{fig4_TB}
\end{figure}

In view of specific experimental protocols, the aspects of the implementation are important and form the key to resolving the main hurdle on the route to observing the rich physics presented by the interplay of crystalline symmetries, fragile topology and non-Abelian band nodes.
Indeed, although implementing the non-trivial winding of three-band Euler class might be more involved than a Chern insulator, optical lattices offer a versatile toolbox to overcome these difficulties, where different orbitals (pseudospin structure) can be encoded as sublattice degrees of freedom (and e.g. optimized in Kagome geometry), or most easily as different internal states of an atom where the tunneling can be engineered via Raman couplings. In addition, optical flux lattices offer a promising platform where a topologically non-trivial Euler model can be engineered in momentum space~\cite{Cooper11_PRL_fluxlatt,Cooper13_PRL_fluxlatt}. Finally, the general nature of the algorithm for generating Euler class models does allow for flexibility in lattice geometry and can be tailored on a case-by-case basis. To illustrate this, by relaxing the condition of flatness of the bands and optimizing for the fewest number of tunneling processes, we arrive at an equally valid Hamiltonian on the square lattice with Euler class $\xi=2$ 
\begin{equation}
\begin{aligned}
    h_3(\boldsymbol{k}) &= 
    -t_a (\cos k_x - \cos k_y)  \;,\\
    h_8 (\boldsymbol{k}) &= 
    t_a \sqrt{3}/2  [2(\cos k_x + \cos k_y) - 3 (\cos 2k_x + \cos 2k_y)]\;,\\
    h_1(\boldsymbol{k}) &= 
     t_b \sin k_x \sin k_y\;, \\
    h_4 (\boldsymbol{k}) &= 
    t_c  \sin 2k_x\;,\\
    h_6 (\boldsymbol{k}) &= 
    t_c  \sin 2k_y\;,
\end{aligned}\label{eq::alternativemodel}
\end{equation}
where we listed the non-zero contributions to Eq. \eqref{eq::App:mainham} for $t_a = 0.35$, $t_b = 0.46$, $t_c = 0.69$. 
In the $(A,B,C)$-orbital basis, the Hamiltonian thus reads
\begin{equation}
\begin{array}{rclrcl}
    h_{AA}&=& h_3 + 1/\sqrt{3} h_8 \,,& h_{AB} &=& h_1\,, \\
    h_{BB}&=& -h_3 + 1/\sqrt{3} h_8 \,, & h_{AC} &=& h_4\,, \\
    h_{CC}&=& -2/\sqrt{3} h_8 \,, & 
    h_{BC} &=& h_6 \,,
\end{array}
\end{equation}
which amounts to the tight-binding parameters
\begin{equation}
\begin{aligned}
    t^{(0,\pm1)}_{AA} &= t^{(\pm1,0)}_{BB} = -t^{(\pm1,0)}_{CC} =- t^{(0,\pm1)}_{CC} =2t_a \,, \\
    t^{(\pm2,0)}_{AA} &= t^{(0,\pm2)}_{AA} = t^{(\pm2,0)}_{BB} = t^{(0,\pm2)}_{BB} = - t^{(\pm2,0)}_{CC}/2 = -t^{(0,\pm2)}_{CC}/2 \\
    &= - t_a  3/2 \,\nonumber, 
    \end{aligned}
\end{equation}
\begin{equation}
\begin{aligned}
    t^{(\pm1,\pm1)}_{AB} &= -t^{(\pm1,\mp1)}_{AB} = -t_b/2\,,\\
    t^{(\pm2,0)}_{AC} &= \pm t_c \,,\\
    t^{(0,\pm2)}_{BC} &= \pm t_c \,.
\end{aligned}
\end{equation}
In the above the super scripts convey the vectors connecting the neighbors of each site and the subscripts indicate the orbitals. These parameters are illustrated in Fig.~\ref{fig4_TB}. We however stress once more that the models can be further fine-tuned to individual setups as the Euler class supersedes any specific crystal structure that accommodates the $C_2\mathcal{T}$ symmetry. 

\section{Quaternions, Rotations and Hopf Fibration}\label{sec:quaternions}
We find that the Hopf map description of quench dynamics can be most conveniently described in terms of quaternions, that directly relate to usual characterizations of rotations in terms of $SU(2)$ and $SO(3)$ matrices, thereby also exposing the intimate relation between the two. Note that here we deal with the first Hopf map, where higher Hopf maps can also be represented by similar constructions, for example using octonions.

\parag{Quaternions}--
Recall that quaternions in essence constitute a generalization of complex numbers in an analogous manner in which the latter extends the real numbers. Specifically, a quaternion $q$ is written as 
\begin{equation}
q=x_0+x_1{ i}+x_2{ j}+x_3{ k},    
\end{equation}
where the quaternion units satisfy ${ij}=k, {jk}=i, {ki}=j$ and ${i}^2={j}^2={k}^2=-1$. We hence see that the real coefficients identify $q$ with $\mathbf{R}^4$ similar to how $\mathbf{C}$ relates to $\mathbf{R}^2$. Accordingly, we refer to $x_0$ and $\{x_1,x_2,x_3\}$ as the real part and pure quaternion part of $q$, respectively. Using the relations between the units and defining the conjugate of $q$ as reversing the sign of the units, that is $q^{*}=-\frac{1}{2}(q+iqi+jqj+kqk)$, 
the norm of $q$ is readily found to be $|q|=\sqrt{qq^{*}}=\sqrt{x_0^2+x_1^2+x_2^2+x_3^2}$.

\parag{Rotations and versors}-- In the subsequent, we will be particularly interested in quaternions $v$ having unit norm, or so-called versors. Due to the constraint $|v|=1$, it is evident that the subset of versors span the hypersphere $S^3\subset\mathbf{R}^4$. Versors are of particular use in describing rotations in three spatial dimensions. Note that we can represent a vector $\mathbf{t}\in\mathbf{R}^3$ as a pure quaternion,  having a zero real part and with the quaternion units corresponding to the unit vectors parametrizing $\mathbf{R}^3$. It is easy to see that upon multiplying a pure quaternion with an arbitrary quaternion results in another pure quaternion. Rotations on vectors can then be implemented using versors, that, due to the unit norm, indeed generate an isometry. In particular, it is well known that acting with a versor $v=x_0+\mathbf{v}$ on a vector $\bf{t}$ as 
\begin{equation}\label{eq::versorrot}
R_v: {\bf t}\mapsto v{\bf{t}} v^{-1},    
\end{equation}
where $v^{-1}$ refers to the inverse of $v$, implements a rotation around the vector $\mathbf{v}$ by an angle $\theta=2\arccos{(x_0)}=2\arcsin{(1-x_0)}$. 
 
\parag{Hopf map}-- The formulation of rotations in terms of versors provides for a direct parametrization of the Hopf map. Indeed, we can reinterpret Eq. \eqref{eq::versorrot} as a map relating the versor $v$ with a vector ${\bf t'}=v{\bf{t}} v^{-1}$. Moreover, as this entails an isometry we can already expect that when ${\bf t}$ is a unit vector, and thus an element corresponding to the two-sphere $S^2$, the map returns another unit vector ${\bf t}'\in S^2$. In other words, the rotation isometry acts on $S^2$ transversely. As the set of versors span $S^3$, this directly induces a map of $S^3$ to $S^2$, the famous Hopf fibration. 

As a specific example one may consider the versor $v=x_0+x_1{ i}+x_2{ j}+x_3{ k}$ and the unit vector in the $\hat{x}$-direction that we thus represent as $i$ in the quaternion language, ${\bf t}=i$. Simple algebra then shows that applying the rotation isometry, Eq. \eqref{eq::versorrot}, results in
\begin{equation}
{\bf t'}=
\begin{pmatrix}
x_0^2+x_1^2-x_2^2-x_3^2\\ 
2x_1x_2+2x_0x_3\\
-2x_0x_2+2x_1x_3
\end{pmatrix},
\end{equation}
which evidently has unit norm, thereby inducing the Hopf map $\mathcal{H}: S^3 \rightarrow S^2$. Indeed, it is easy to verify that the inverse image is a circle, which constitutes the familiar fibre under this map.

We can make the above even more insightful upon representing the action of the versor $v$ in terms of a rotation matrix acting on a vector. The columns represent the action on the unit vectors $i,j,k$ and a similar calculation to the one above then gives us the general rotation matrix 
\begin{widetext}
\begin{equation}\label{eq::rotationversorrep}
R_v=
\begin{pmatrix}
x_0^2+x_1^2-x_2^2-x_3^2 & 2x_1x_2-2x_0x_3 & 2x_0x_2+2x_1x_3 \\
2x_1x_2+2x_0x_3 & x_0^2-x_1^2+x_2^2-x_3^2& -2x_0x_1+2x_2x_3\\
-2x_0x_2+2x_1x_3 & 2x_0x_1+2x_2x_3 & x_0^2-x_1^2-x_2^2+x_3^2 
\end{pmatrix}.  
\end{equation}
\end{widetext}
It is routinely verified that the columns and rows of $R_v$ are orthogonal and that $R_v$ indeed represents the familiar rotation matrix in 3D. From the viewpoint of the Hopf map, we note that Eq.~\eqref{eq::rotationversorrep} is nothing but an explicit parameterization. Indeed, any row, column or linear linear combination generated by acting on $R$ with a unit norm vector from the left or right implements the first Hopf map.

\parag{Quenches in two-band systems}-- We can directly put the above notions in use to reproduce the dynamics of two band models quenched between trivial and non-trivial Chern numbers \cite{Wangchern_2017, Tarnowski19_NatCom, ChernHopf_Yu, YiPan19_arx_hopfTori,ChernHopf_Yu}. We consider the model given in Eq.\eqref{appeq::2band} for $d_0=0$,
where the relation of $\bf{ d}$ to the Chern number is given in the main text Eq.~\textcolor{blue}{(2)}.
An essential role is then played by the time evolution operator $U_{\mathcal{C}}$, which takes the particularly easy form 
\vspace{-.5mm}
\begin{equation}
U_{\mathcal{C}}=e^{-itH_{\mathcal{C}}}=\cos{(t)}-i\sin{(t)}\boldsymbol{\sigma}\cdot \bf{ d}(\bf{k}).
\end{equation}
Writing this in matrix form one obtains
\begin{equation}
U_{\mathcal{C}}=
\begin{pmatrix}
x_0+i x_3 & ix_1 + x_2\\
i x_1 - x_2 & x_0 - i x_3\\
\end{pmatrix},  
\end{equation}
where we already suggestively identified the elements $\{x_0,x_1,x_2,x_3\}=\{\cos{t}, -\sin(t)d_1, -\sin(t)d_2, -\sin(t)d_3\}$ in terms of the components $d_i(\bf{k})$ of ${\bf d}(\bf{k})$. Indeed, assuming that upon spectral flattening ${\bf d}(\bf{k})$ is a unit vector, we can directly see that, by relating the above to a versor with components $x_\alpha$, the action of the time evolution operator on the initial state $\Psi_0(\bf{k})$ is simply to induce a rotation around the vector ${\bf d}(\bf{k})$ with a period that is set by $|\bs{d}|$. In fact, the above identification is the standard one to relate a quaternion of unit norm to a $SU(2)$ matrix representation. 

Starting from a trivial $\Psi_0(\bf{k})$, we quench the system suddenly with a non-trivial Hamiltonian so that the state evolves as $\Psi({\bs k},t)=U_\mathcal{C}\Psi_0(\bf{k})$. One can then map the evolving state back to the Bloch sphere upon considering $\hat{p}=(\Psi^{\dagger}({\bs k},t)\sigma_x\Psi({\bs k},t),\Psi^{\dagger}({\bs k},t)\sigma_y\Psi({\bs k},t),\Psi^{\dagger}({\bs k},t)\sigma_z\Psi({\bs k},t))^{\top}$, thereby establishing a Hopf map from $\{k_x,k_y,t\}$, that is identified with $S^3$, to the Bloch sphere constituting the $S^2$ \cite{Wangchern_2017}. 
This can be directly checked from the above formulae by taking, e.g.~$\Psi_0=(1,0)^{\top}$, which gives
\begin{equation}
\hat{p}=
\begin{pmatrix}
2x_1x_3-2x_0x_2\\
2x_0x_1+2x_2x_3\\
x_0^2+x_3^2-x_1^2-x_2^2
\end{pmatrix},
\end{equation}
thus parameterizing the Hopf map with the last row (or column upon taking a minus sign in the definition of $x_\alpha$) of Eq.~\eqref{eq::rotationversorrep}. Similarly, one can verify that taking $\Psi_0({\bs k})=(0,1)^{\top}$ results in a similar expression. This, therefore, shows that starting from an arbitrary normalized initial state, the first Hopf map is realized with the construction of Eq.~\eqref{eq::rotationversorrep}. Most interestingly, the Hopf map directly exposes the non-trivial Chern number of the quench Hamiltonian as it implies a non-trivial Hopf invariant $\cal{H}$~\cite{Wangchern_2017,ChernHopf_Yu}, which is also manifested in non-trivial linking of the trajectories under the inverse image of $\cal{H}$  that can be measured in experiments \cite{WeiPan18_PRL_hopfExp,Tarnowski19_NatCom, YiPan19_arx_hopfTori}. We note that, the quaternion description also naturally captures quenches from topologically non-trivial to trivial Hamiltonians.

\section{Quench dynamics in three-band models having non-trivial Euler class}\label{sec:quench}
We now turn our attention to further detailing the main subject, topological aspects and dynamics of quenches involving non-trivial Euler class Hamiltonian.  We thus assume $C_2{\cal T}$-symmetry in all instances, which in fact is rather rudimentary and hence does not impose a serious limitation with regard to experimental implementation. 

\parag{Quenching with non-trivial Chern Hamiltonian}--
We first consider quenching an initial state with a Hamiltonian having a non-trivial Euler class in which the third band $|n({\bf k})\rangle$ has the highest energy, topping the other two degenerate bands by an energy gap. We then return to the inverted situation, with $| n({\bf k})\rangle$ being the bottom band, at the end of this section.

As shown in the main text, upon spectral flattening, the Hamiltonian takes the form of Eq.\eqref{eq::specflat;app}. A non-trivial Euler class, in analogy to the Chern number, is then manifested by the geometrical interpretation that $n(\bs k)$ traces the unit sphere, in fact by a factor 2 as compared to the Chern number case. Moreover, we see that Hamiltonian \eqref{eq::specflat;app} physically represents a rotation by angle $\pi$ around the vector $n(\bs k)$. Indeed, the quaternion description in this case reads $v=\cos(\pi/2)+\sin(\pi/2)\{n_1i+n_2j+n_3k\}$ in terms of the components of $n(\bs k)=(n_1,n_2,n_3)^\top$. As a result, the time evolution operator $U$ still assumes the simple form,
\begin{equation}
U=e^{-itH}=\cos{(t)}-i\sin{(t)}H(\bs{k}),
\end{equation}
reminiscent of the two-band case. 

Starting with a trivial normalized state $\Psi_0(\bs{k})$ we then want to appeal to the above Hopf construction. In this regard we first focus on the case $\Psi_0({\bs k})=(1,0,0)^{\top}$. Applying $U$ on $\Psi_0({\bs k})$ we obtain
\begin{equation}
\Psi({\bs k},t)=
\begin{pmatrix}
\cos(t)-i\sin(t)(2n_1^2-1)\\
-i\sin(t)2n_1n_2\\
-i\sin(t)2n_1n_3\\
\end{pmatrix}.  
\end{equation}
A priori, it seems hard to relate to the aforementioned Hopf construction. However, recall that we are free to reparametrize the Hopf map upon applying rotations to the initial vector ${\bf t}$ as in Eq. \eqref{eq::rotationversorrep}. Additionally, $H$ is nothing but a rotation of  $\Psi_0({\bs k})$ around $\bs n(\bs k)$. 
We therefore relabel $(2n_1^2-1, 2n_1n_2, 2n_1n_3)$ as $(a_1,a_2,a_3)$, which evidently sill amounts to a unit vector.  Consequently, we thus find the following form of the time-evolved state 
\begin{equation}\label{eq:::app::evolvpsi}
\Psi({\bs k},t)=
\begin{pmatrix}
\cos(t)-i\sin(t)a_1\\
-i\sin(t)a_2\\
-i\sin(t)a_3\\
\end{pmatrix}.  
\end{equation}
With the above Eq. \ref{eq:::app::evolvpsi} in hand, we subsequently define 
\begin{widetext}
\begin{equation}
 \mu_x=
 \begin{pmatrix}
0 &  i & 1\\
-i & 0 & 0\\
1 & 0 & 0
\end{pmatrix},
\qquad
\mu_y=
 \begin{pmatrix}
0 &  1 & -i\\
1 & 0 & 0\\
i & 0 & 0
\end{pmatrix},
\qquad
\mu_z=
\begin{pmatrix}
1 &  0 & 0\\
0 & -1 & 0\\
0 & 0 & -1
\end{pmatrix}.
\end{equation}
\end{widetext}
These matrices then project $\Psi(t)$ back to a `Bloch vector' $\hat{p}\in S^2$ in the desired manner, upon contracting 
\begin{equation}\label{eq::Bloch vector}
\hat{p}=(\Psi^{\dagger}(t)\mu_x\Psi(t), \Psi^{\dagger}(t)\mu_y\Psi(t), \Psi^{\dagger}(t)\mu_z\Psi(t))^{\top}.
\end{equation}
Indeed, identifying $\{x_0,x_1,x_2,x_3\}=\{\cos{t}, -\sin(t)a_1, -\sin(t)a_2, -\sin(t)a_3\}$, we observe that this parametrizes the first Hopf map by the first column of Eq. \eqref{eq::rotationversorrep}.

A few remarks on the above are in place. First, we note that multiplying the above matrices with rotations (by an angle $\pi/2$) that interchange the bottom two components merely interchanges the components of the Hopf parameterization, thus preserving the map.
Secondly, we observe a close analogy to the two-band case. A quaternion can alternatively be rewritten as two complex numbers. That is, we can define $q=z_1+z_2j$ in terms of $z_1=x_0+ix_1$ and $z_2=-x_2+ix_3$.  Interpreting these numbers as a vector $\zeta=(z_1,z_2)^{\top}$, we obtain the same expression for the Hopf map upon replacing the ${\bs \mu}$-matrices in Eq. \eqref{eq::Bloch vector} with the respective standard Pauli matrices ${\bs \sigma}$ and the three vector $\Psi({\bs k},t)$ with the two vector $\zeta({\bs k},t)$. Thirdly, on a related note, we see that these identifications can similarly be established for the other basis vectors taken as the initial state $\Psi_0({\bs k})=(0,1,0)^{\top}$ and $\Psi_0({\bs k})=(0,0,1)^{\top}$.
For these choices we find respectively,
\begin{equation}
\Psi({\bs k},t)=
\begin{pmatrix}
-i\sin(t)b_1\\
\cos(t)-i\sin(t)b_2\\
-i\sin(t)b_3\\
\end{pmatrix},  
\end{equation}
and 
\begin{equation}
\Psi({\bs k},t)=
\begin{pmatrix}
-i\sin(t)c_1\\
-i\sin(t)c_2\\
\cos(t)-i\sin(t)c_3\\
\end{pmatrix}, 
\end{equation}
where $(b_1,b_2,b_3)=(2n_2n_1, 2n_2^2-1, 2n_2n_3)$ and $(b_1,b_2,b_3)=(2n_3n_1, 2n_2n_3, 2n_3^2-1)$. Evidently, the Hopf map can then be parametrized in an analogous manner. We nonetheless need to take into account that the two purely imaginary components, constituting the second complex number when written as the two vector $\zeta({\bs k},t)$, are shuffled. In other words, for $\Psi_0({\bs k})=(0,1,0)^{\top}$ or $\Psi_0({\bs k})=(0,0,1)^{\top}$, we need to respectively replace $\mu_i\mapsto u\mu_i u$ or $\mu_i\mapsto v\mu_i v$, where
\begin{equation}
u=
\begin{pmatrix}
0 &  1 & 0\\
1 & 0 & 0\\
0 & 0 & 1
\end{pmatrix}, \text{ and } \:
v=
\begin{pmatrix}
0 &  0 & 1\\
0 & 1 & 0\\
1 & 0 & 0
\end{pmatrix}. 
\end{equation}
The general parametrization for any normalized initial state can then similarly be achieved, in terms of the general $\pi$-rotated vector ${\bs a}({\bs k})$.

Upon the relabelling in terms of the $\bs{a,b,c}$-vectors, the properties and consequences of the outlined Hopf map are thus determined by the inner topological structure, irrespective of the initial state (cf.~Fig.\ref{fig:App2_tomography}). That is, it amounts to a $\pi$- rotated vector that wraps and unwraps the sphere when ${\bs n}({\bs k})$ traces the sphere once, as detailed in the main text. 

Finally, let us close this section by commenting on the inverted models, in which the third non-degenerate band is at the bottom of the spectrum. In this case the flattened Hamiltonian is of the form,
\begin{equation}\label{eq::specflatINV;app}
H(\bs{k}) = -2\, \bs{n}(\bs{k})\cdot  \bs{n}(\bs{k})^\top+\mathbb{I}_3.
\end{equation} 
Repeating the procedure above, we notice that the effect of this change in Hamiltonian is to induce an extra minus sign in the $x_1,x_2$ and $x_3$ components. Hence, we observe that this merely changes the parametrization of the Hopf map by interchanging the respective row with a column in Eq. \eqref{eq::rotationversorrep}.

\section{Tomography in 3-band Euler model}\label{App:Tomography}
\begin{figure}
	\centering\includegraphics[width=.75\linewidth]{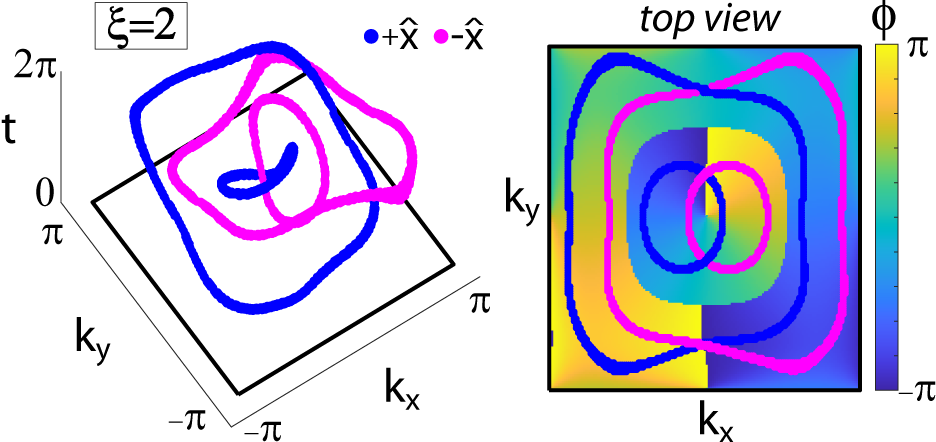}
	\caption{ Linking of the inverse images of $\pm\hat{x}$ for initial state $\Psi_0=(0,0,1)^\top$, where the Bloch vector $\bs{\hat{p}}$ is reconstructed from the momentum distribution via tomography.	}
	\label{fig:App2_tomography}
\end{figure}

As explained in the main text, a state tomography can be employed to measure the linking in $(k_x,k_y,t)$-space. We here present the linking structure acquired through the proposed tomography scheme, for another initial state $\Psi_0=(0,0,1)^\top$ to simultaneously illustrate the effect of different initial states. Our tomography scheme involves first applying a $\pi$-pulse with respect to sublattice $B$ to access the two-vector [depending on the initial state, here $z_1$ is identified with the last component of $\Psi(\bs{k},t)$ as it involves the real $x_0$-term, $\zeta=(\Psi_3,\Psi_1+i\Psi_2)^\top$]. Secondly, we quench with the tomography Hamiltonian having flat bands with respect to sublattice $C$, since we start with an initial state completely localized in $C$. Namely, $\mu'_z=v\mu_zv=diag(-1,-1,1)$ and $H_{tom}=(\omega/2)\mu'_z$. After evolving with $H_{tom}$ for a time $t_{tom}$, we obtain the momentum distribution (which can be measured in TOF) as $m(\bs{k},t_{tom})\propto(1+\sin\theta_{\bs{k}}\cos(\phi_{\bs{k}}+\omega t_{tom}))$. The azimuthal angle $\phi_{\bs{k}}$ and the amplitude $\sin\theta_{\bs{k}}$ is calculated by fitting $m(\bs{k},t_{tom})$ with a cosine at each $\bs{k}$ and quench time $t$, for which the results are given in Fig.~\ref{fig:App2_tomography}.

Note that, for the initial state $(0,0,1)$ the monopole--antimonopole pair now resides in patches aligned at the center and outer circle of the BZ as can be seen in Fig.~\ref{fig:App2_tomography}, as opposed to left/right division visible in Fig.~\textcolor{blue}{1} and Fig.~\textcolor{blue}{2} given for $(1,0,0)$. 
This can also be seen from the stereo-graphic projection of the components of $H\Psi_0({\bs k})$.  We emphasize that despite the different alignments of the BZ patches, the clear separation can be easily seen in the azimuthal angle profile given in Fig.~\ref{fig:App2_tomography}b, and the monopole--antimopopole pair is topologically stable.

\section{String and monopole charges on the real projective plain}\label{App:TopAsp}
We recall some aspects of the real projective plane. Turning first to the Euler Hamiltonian, we note that the models have a gauge symmetry, relating each eigenstate spanning the dreibein  $E=\{\ket{u_1(\bs k)},\ket{u_2(\bs k)},\ket{u_3(\bs k)}\}$ with its negative partner. This therefore relates to the real protective plane ${\bs R \bs P^2}$, as in a biaxial nematic \cite{Nissinen2016, Beekman20171, volovik2018investigation, Kamienrmp, Prx2016}. The string charges corresponding to first homotopgy group $\pi_1$ are thus characterized by $\pi_1({\bs R \bs P^2}^2)={\bs Z}_2$, whereas the monopole charges are given by the second homotopy group $\pi_2({\bs R \bs P}^2)={\bs Z}$. The non-triviality of the string charges dictates caution in considering maps from the torus by using intuition from the sphere. Their presence ensures the existence of weak invariants.  In each of the two directions the weak index can take either a trivial or non-trivial value, giving four possibilities. This effects the monople charge possibilities. Namely, when there are nonzero weak invariants in either direction or both, an even multiple of monopole charges can adiabatically be split in two equal pieces and transferred around the non-trivial direction. The action of $\pi_1$ then ensures that the charge gets opposite values and thus can annihilate the other half, rendering a ${\bs Z}_2$ classification. 
We therefore restrict attention to systems having trivial string charges. In fact the models presented are designed to meet this criterion. However even in this case, the direction is not defined and hence skyrmions and anti-skyrmions can not be discriminated in {\it an absolute sense}, showing that these charges are characterized by the absolute value of the winding number. This also relates to Alice dynamics, as in certain scenarios the monopole can be adiabatically deformed into an Alice string, which changes the sign of the charge charge upon passing through \cite{AlicestringVolovik,Schwarz_alice}. Details of these features are beyond the scope of this paper and will be reported elsewhere. To construct Euler class, an orientation must be fixed, which physically amounts to specifying a handiness. Once this has been taken into account the sphere analogy becomes appropriate.

\section{Specific matrix parametrization of the models}\label{App:matrices Hamiltonian}
We here give the explicit forms of the tunneling matrix elements given in Eq. \eqref{eq::modelparameters}.

The Euler Hamiltonian with $\xi=2$ can be constructed by restricting $N=2$-neighbor tunneling~\cite{newpaperfragile}. The eight $t_j(\alpha, \beta)$ multiplying the Gell-Mann matrices are $5\times 5$ matrices given as:
\begin{widetext}
\begin{equation}
t_1=
 \begin{pmatrix}
 0.0089 - 0.0151i & -0.0761 + 0.0309i & -0.0025 - 0.0076i &  0.0811 - 0.0158i & -0.0139 + 0.0000i\\
 -0.0761 + 0.0309i & -0.1205 - 0.0467i &  0.0025 + 0.0233i &  0.1155 + 0.0000i &  0.0811 + 0.0158i\\
 -0.0025 - 0.0076i &  0.0025 + 0.0233i & -0.0025 + 0.0000i  & 0.0025 - 0.0233i & -0.0025 + 0.0076i\\
 0.0811 - 0.0158i  & 0.1155 + 0.0000i  & 0.0025 - 0.0233i & -0.1205 + 0.0467i & -0.0761 - 0.0309i\\
  -0.0139 + 0.0000i  & 0.0811 + 0.0158i & -0.0025 + 0.0076i & -0.0761 - 0.0309i &  0.0089 + 0.0151i
\end{pmatrix}
\end{equation}
\begin{equation*}
t_3=
 \begin{pmatrix}
   -0.0025 &  -0.0883&   -0.1727&   -0.0883&   -0.0025\\
    0.0833  & -0.0025 &   0.0375 &  -0.0025 &   0.0833\\
    0.1677   &-0.0425  & -0.0025  & -0.0425  &  0.1677\\
    0.0833  & -0.0025   & 0.0375   &-0.0025   & 0.0833\\
   -0.0025   &-0.0883   &-0.1727   &-0.0883   &-0.0025
\end{pmatrix},
\quad t_4=
 \begin{pmatrix}
  0.0278i&    0.2275i&    0.4917i&    0.2275i&    0.0278i\\
  0.0636i &   0.1142i &   - 0.2810i&    0.1142i&   0.0636i\\
  0 &0& 0& 0& 0\\
  -0.0636i &  - 0.1142i&    0.2810i &  - 0.1142i & - 0.0636i\\
  -0.0278i  & - 0.2275i &  - 0.4917i &  - 0.2275i & - 0.0278i\\
\end{pmatrix}
\end{equation*}
\begin{equation*}
t_6=
 \begin{pmatrix}
 0.0278i&   0.0636i&    0&   - 0.0636i&   - 0.0278i\\
 0.2275i &  0.1142i &   0 & - 0.1142i  & - 0.2275i\\
 0.4917i  & - 0.2810i&  0  &   0.2810i  & - 0.4917i\\
 0.2275i   & 0.1142i  & 0   &- 0.1142i   &- 0.2275i\\
 0.0278i    &0.0636i   &0    &- 0.0636i   &- 0.0278i
\end{pmatrix},
\quad t_8=
 \begin{pmatrix}
    0&      -0.1879&   -0.4330&   -0.1879&         0\\
   -0.1879&    0&       0.3083 &   0      &   -0.1879\\
   -0.4330 &   0.3083   & 0     &0.3083    &  -0.4330\\
   -0.1879  &    0 &    0.3083   & 0        &-0.1879\\
    0 &     -0.1879   & -0.4330 & -0.1879    &  0
\end{pmatrix},
\end{equation*}
whereas $t_2=t_5=t_7=0$ as they correspond to the complex Gell-Mann matrices.

Similarly, the Euler Hamiltonian with $\xi=4$ can be constructed by restricting $N=3$-neighbor tunneling. Accordingly, $t_j(\alpha, \beta)$ are $7\times 7$ matrices given by:
\begin{multline*}
t_1=\\
 \!\!\!\!\!\!\!\!\!\!\!\!
 \begin{pmatrix}
   0             &      0.0626 - 0.0037i&   0.1137 - 0.0240i&    - 0.0001i&     -0.1137 + 0.0242i&       -0.0626 + 0.0036i&  0.0001i\\
  -0.0626 + 0.0037i&   0  &   0.0421 + 0.0278i&    - 0.0037i&    -0.0421 - 0.0204i &      - 0.0073i&        0.0626 + 0.0036i\\
  -0.1137 + 0.0240i & -0.0421 - 0.0278i   &  0                &   - 0.0241i &       0.0482i        &  0.0421 - 0.0204i&    0.1137 + 0.0242i\\
   0.0001i           & 0.0037i             &0.0241i            &       0     &      - 0.0241i       &    - 0.0037i&       - 0.0001i\\
   0.1137 - 0.0242i   &0.0421 + 0.0204i     &- 0.0482i          &  0.0241i    &         0           &-0.0421 + 0.0278i&   -0.1137 - 0.0240i\\ 
   0.0626 - 0.0036i    &0.0073i        &   -0.0421 + 0.0204i    &0.0037i       &     0.0421 - 0.0278i&       0&           -0.0626 - 0.0037i\\
   0.0001i           &-0.0626 - 0.0036i &  -0.1137 - 0.0242i     &  0           &0.1137 + 0.0240i     &  0.0626 + 0.0037i&  0
\end{pmatrix}
\end{multline*}
\begin{equation*}
t_3=
 \begin{pmatrix}
   -0.0298&   -0.0125&    0.0432&    0.0432&    0.0432&   -0.0125&   -0.0298\\
   -0.0125 &  -0.1652 &  -0.0550 &   0.0753 &  -0.0550 &  -0.1652 &  -0.0125\\
    0.0432  & -0.0550  &  0.1086  & -0.0601  &  0.1086  & -0.0550  &  0.0432\\
    0.0432   & 0.0753   &-0.0601   &-0.0073   &-0.0601   & 0.0753   & 0.0432\\
    0.0432&   -0.0550&    0.1086&   -0.0601&    0.1086&   -0.0550&    0.0432\\
   -0.0125 &  -0.1652 &  -0.0550 &   0.0753 &  -0.0550 &  -0.1652 &  -0.0125\\
   -0.0298  & -0.0125  &  0.0432  &  0.0432  &  0.0432  & -0.0125  & -0.0298
\end{pmatrix}
\end{equation*}
\begin{equation*}
\!\!\!\!\!\!\!\!\!\!\!\!\!\!\!\!\!\!\!\!\!
t_4\!\!=\!\!\!
 \begin{pmatrix}\!
        0&   -0.0531&   -0.1480&   -0.1995&   -0.1480&   -0.0531&         0\\
    0.0531&         0&   -0.0655&   -0.0840&   -0.0655&         0&    0.0531\\
    0.1480 &   0.0655 &        0 &   0.3123 &        0 &   0.0655 &   0.1480\\
    0.1995  &  0.0840  & -0.3123  &       0  & -0.3123  &  0.0840  &  0.1995\\
    0.1480   & 0.0655   &      0   & 0.3123   &      0   & 0.0655   & 0.1480\\
    0.0531    &     0  & -0.0655  & -0.0840  & -0.0655    &     0    &0.0531\\
         0   &-0.0531   &-0.1480   &-0.1995   &-0.1480   &-0.0531     &    0
\end{pmatrix}
\! t_6\!\!=\!\!\!\!
 \begin{pmatrix}
  -0.0508&   -0.0735&    0.0490&         0&   -0.0490&    0.0735&    0.0508\\
   -0.0735&   -0.1579&   -0.2956&         0&    0.2956&    0.1579&    0.0735\\
    0.0490 &  -0.2956 &   0.2493 &        0 &  -0.2493 &   0.2956 &  -0.0490\\
         0  &       0  &       0  &       0  &       0  &       0  &       0\\
   -0.0490   & 0.2956   &-0.2493   &      0   & 0.2493   &-0.2956   & 0.0490\\
    0.0735    &0.1579    &0.2956    &     0   &-0.2956   &-0.1579   &-0.0735\\
    0.0508    &0.0735  & -0.0490     &    0    &0.0490   &-0.0735   &-0.0508
\end{pmatrix}
\end{equation*}
\begin{equation}
t_8=
 \begin{pmatrix}
        0&   -0.0461&         0&    0.1148&         0&   -0.0461&         0\\
   -0.0461&         0&   -0.1879&   -0.4330&   -0.1879&         0&   -0.0461\\
         0 &  -0.1879 &        0 &   0.3083 &        0 &  -0.1879 &        0\\
    0.1148  & -0.4330  &  0.3083  &       0  &  0.3083  & -0.4330  &  0.1148\\
         0   &-0.1879   &      0   & 0.3083   &      0   &-0.1879   &      0\\
   -0.0461    &     0   &-0.1879   &-0.4330   &-0.1879    &     0   &-0.0461\\
         0   &-0.0461    &     0    &0.1148    &     0   &-0.0461    &     0
\end{pmatrix}.
\end{equation}
\end{widetext}


\end{document}